\documentclass[aps,prd,preprintnumbers,superscriptaddress,nofootinbib,notitlepage,twocolumn,floatfix]{revtex4-2}
\usepackage[pdftex]{graphicx}
\usepackage{bm,latexsym,amsmath,amssymb,amsfonts,mathrsfs}
\usepackage{color}
\allowdisplaybreaks[1]
\usepackage[pdftex,colorlinks=true,linkcolor=blue,citecolor=cyan]{hyperref}
\newcommand*{\D}{\mathrm{d}}

\begin{document}
\title{Distinguishing modified gravity with just two tensorial degrees of freedom from general relativity: Black holes, cosmology, and matter coupling}
%
\author{Aya~Iyonaga}
\email[Email: ]{iyonaga@rikkyo.ac.jp}
\affiliation{Department of Physics, Rikkyo University, Toshima, Tokyo 171-8501, Japan
}
\author{Tsutomu~Kobayashi}
\email[Email: ]{tsutomu@rikkyo.ac.jp}
\affiliation{Department of Physics, Rikkyo University, Toshima, Tokyo 171-8501, Japan
}
%
\begin{abstract}
We consider spatially covariant modified gravity in which the would-be
scalar degree of freedom
is made non-dynamical and hence there are just two tensorial degrees of freedom,
i.e., the same number of dynamical degrees of freedom as in general relativity.
Focusing on a class of such modified gravity theories characterized by
three functions of time, we discuss how modified gravity with two tensorial
degrees of freedom can be distinguished observationally or phenomenologically
from general relativity. It is checked that the theory gives the same predictions as general
relativity for weak gravitational fields and the propagation speed of gravitational waves.
We also find that there is no modification to asymptotically flat black hole solutions.
Due to a large degree of freedom to choose the time-dependent functions in the theory,
the homogeneous and isotropic cosmological dynamics can be made close to or
even identical to that of the $\Lambda$CDM model. We investigate
the behavior of cosmological perturbations in the long and short wavelength limits
and show that in both limits the effects of modified gravity appear only through
the modification of the background evolution. Finally, it is remarked that
in the presence of a galileon field in the matter sector,
the scalar degree of freedom is revived, ruining the essential feature of the theory.
\end{abstract}
\preprint{RUP-21-16}
\maketitle
\section{Introduction}

According to Lovelock's theorem~\cite{Lovelock:1971yv,Lovelock:1972vz},
the Einstein tensor plus a cosmological term
is the only possible second-order Euler-Lagrange equation that is
obtained from a diffeomorphism invariant action
constructed from the metric tensor alone in four spacetime dimensions.
Modifying gravity amounts to relaxing the postulates of the theorem.
Abandoning full spacetime diffeomorphism invariance, one can for example consider
modified gravity enjoying only three-dimensional spatial diffeomorphism invariance.
Basically, this way of modifying gravity
is equivalent to introducing a new dynamical scalar degree of freedom (DOF)
that spontaneously breaks time diffeomorphism invariance.
This is the underlying idea utilized in the construction of
effective field theories for inflation and dark
energy~\cite{Cheung:2007st,Gubitosi:2012hu,Bloomfield:2012ff,Gleyzes:2014rba}.
A general framework for scalar-tensor theories based on this idea
has been developed in Refs.~\cite{Gao:2014soa,Gao:2014fra}.

 An interesting twist is that a scalar DOF incorporated in such a framework
 is not necessarily dynamical and it is possible that
 modified gravity in which time diffeomorphism invariance is broken
 has only two tensorial degrees of freedom as general relativity (GR).
 The simplest example of such ``scalarless''
 scalar-tensor theories is the cuscuton theory~\cite{Afshordi:2006ad},
 in which the scalar field has an infinite propagation speed.
An attempt to identify the cuscuton-like subclass within
the Horndeski~\cite{Horndeski:1974wa} and beyond Horndeski~\cite{Gleyzes:2014dya} theories
is presented in Ref.~\cite{Iyonaga:2018vnu}.
Other modified gravity theories with two DOFs have been constructed in
Refs.~\cite{Lin:2017oow,Aoki:2018zcv,Aoki:2018brq,DeFelice:2020eju,Aoki:2020lig},
a subclass of which is in fact equivalent to the cuscuton theory~\cite{Aoki:2021zuy}.
On the basis of the general framework of spatially covariant theories of
gravity~\cite{Gao:2014soa,Gao:2014fra}, Ref.~\cite{Gao:2019twq} has derived
through a rigorous Hamiltonian analysis
the conditions under which there are only two DOFs and presented
a particular action satisfying the conditions.
The action is composed of two derivative terms built out of
three-dimensional geometrical quantities as in GR, but depends
on the lapse function in a nontrivial manner.
A complementary perturbative approach to spatially
covariant modified gravity without a scalar DOF
has been proposed in Ref.~\cite{Hu:2021yaq}.

The purpose of this paper is to explore phenomenological aspects
of the theory of modified gravity with two tensorial DOFs proposed in Ref.~\cite{Gao:2019twq}
to see how the theory can be distinguished observationally from GR.
We will see that it is quite difficult to find differences between
the two theories in the weak gravity regime,
gravitational wave propagation,
black-hole spacetimes, and cosmology,
when restricted to a certain subset of the theories of Ref.~\cite{Gao:2019twq}.
The theory of modified gravity we study in this paper is
somewhat similar to the infrared limit of Ho\v{r}ava gravity~\cite{Horava:2009uw}
in some respects, and we share the same motivation, e.g., as Ref.~\cite{Franchini:2021bpt}.

The structure of this paper is as follows.
In the next section, we review modified gravity theories
with two tensorial DOFs~\cite{Gao:2019twq} which we study in this paper.
In Sec.~\ref{sec:GWs-Newton}, we identify the subclass of the theories
that evades solar-system and gravitational-wave constraints.
Then, we study black hole spacetimes in Sec.~\ref{sec:black-hole} and
cosmological aspects in Sec.~\ref{sec:cosmology}.
In Sec.~\ref{sec:matter}, a side remark concerning a coupling to
a galileon field is given.
Finally, we draw our conclusions in Sec.~\ref{sec:conclusions}.

\section{Spatially covariant gravity with two tensorial degrees of freedom}

In this section,
we introduce a class of spatially covariant modified gravity
in which there are just two tensorial DOFs~\cite{Gao:2019twq}
and in particular there is no scalar DOF.

\subsection{``Scalarless'' scalar-tensor theories}

Let us begin with scalar-tensor theories with
one scalar DOF, $\phi$, on top of two tensorial DOFs.
In the unitary gauge where $\phi$ is homogeneous on constant time hypersurfaces,
ghost-free scalar-tensor theories can in general be described by the action of the
form~\cite{Gao:2014soa,Gao:2014fra},
\begin{align}
S=\int\D t\D^3x\,N\sqrt{\gamma}{\cal L}(N,\gamma_{ij}, K_{ij},R_{ij},D_i;t),
\label{action:general-s-t}
\end{align}
where $N$ is the lapse function, $\gamma_{ij}$ is the spatial metric,
$K_{ij}$ is the extrinsic curvature of constant time hypersurfaces,
\begin{align}
K_{ij}=\frac{1}{2N}\left(\partial_t\gamma_{ij}-D_iN_j-D_jN_i\right),
\end{align}
with $N_i$ being the shift vector, $R_{ij}$ is the intrinsic curvature tensor,
and $D_i$ is the covariant derivative with respect to $\gamma_{ij}$.
The Horndeski action~\cite{Horndeski:1974wa,Deffayet:2011gz,Kobayashi:2011nu}
can be recast in the above form
in terms of these Arnowitt-Deser-Misner (ADM) variables.
One may also include the velocity of the lapse function,
$\partial_t N- N^iD_iN$, and explore a larger class of
healthy scalar-tensor theories with three DOFs~\cite{Gao:2018znj}.
Although such a generalization leads to an interesting class of
theories called degenerate higher-order scalar-tensor
theories~\cite{Langlois:2015cwa,Crisostomi:2016czh,BenAchour:2016fzp},
we focus for simplicity on the action of the form~\eqref{action:general-s-t}.

Thus, in general, the action~\eqref{action:general-s-t} describes
scalar-tensor theories with three DOFs.
However, in a particular subset of theories,
the scalar mode obeys an elliptic equation and does not propagate.
Consequently, such theories have only two tensorial DOFs
as in GR, but the action is certainly different from that of GR.
A particular example is given by~\cite{Gao:2019twq}
\begin{align}
S&=\frac{1}{2}\int\D t\D^3x\, N\sqrt{\gamma}\biggl[
\frac{\beta_0 N}{\beta_2+N}K_{ij}K^{ij}
\notag \\ &\quad\quad
-\frac{\beta_0}{3}
\left(\frac{2N}{\beta_1+N}+\frac{N}{\beta_2+N}\right)K^2
\notag \\ & \quad\quad
+\alpha_1+\alpha_2R
+\frac{1}{N}\left(\alpha_3+\alpha_4R\right)
\biggr],\label{action:Gao}
\end{align}
where $\beta_A$ and $\alpha_A$ are functions of $t$.
A rigorous Hamiltonian analysis has shown that
this theory indeed has only two tensorial DOFs~\cite{Gao:2019twq}.\footnote{The
case with $\beta_0=1$ is presented in Ref.~\cite{Gao:2019twq}, but it
is clear from the footnote of Ref.~\cite{Gao:2019twq} that one can
include this time-dependent coefficient.}
The cuscuton theory~\cite{Afshordi:2006ad} is reproduced
by setting $\beta_0=\alpha_2=1$ and $\beta_1=\beta_2=\alpha_4=0$,
with $\alpha_3$ coming from the cuscuton term
$\sigma(\phi)\sqrt{-\partial_\mu\phi\partial^\mu\phi}$,
while the extended cuscuton theories~\cite{Iyonaga:2018vnu,Iyonaga:2020bmm}
correspond to the case with $\beta_1=\beta_2$.
The covariant expression for the action~\eqref{action:Gao}
in terms of a St\"uckelberg scalar field is given in Appendix~\ref{app:covariantize}.
It is instructive to rewrite the kinetic part of the Lagrangian as
\begin{align}
    \frac{\beta_0}{2}\left(
    \frac{N}{\beta_2+N}\widetilde K_{ij}\widetilde K^{ij}
    -\frac{2}{3}\frac{ N}{\beta_1+N}K^2
    \right),
\end{align}
where $\widetilde{K}_{ij}:=K_{ij}-(1/3)K\gamma_{ij}$ is the traceless part
of the extrinsic curvature. This expression shows the roles of the functions 
$\beta_1$ and $\beta_2$ more clearly.
In the following we will explore various aspects of the gravitational theory
with the action~\eqref{action:Gao}.

Before proceeding let us comment on other examples of ``scalarless''
scalar-tensor theories with two DOFs.
Under the assumption that the Lagrangian is linear in the lapse function
(i.e., ${\cal L}$ in Eq.~\eqref{action:general-s-t} is independent of $N$),
the conditions for a theory to have only two tensorial DOFs
have been addressed in Ref.~\cite{Lin:2017oow}.
References~\cite{Aoki:2018zcv,Aoki:2018brq,DeFelice:2020eju}
obtained gravitational theories with two DOFs
via canonical transformations from GR
in the Hamiltonian formulation.
In the context of the recent claim concerning the $D\to 4$ limit
of $D$-dimensional Gauss-Bonnet gravity~\cite{Glavan:2019inb},
a consistent theory with two DOFs has been proposed~\cite{Aoki:2020lig}
that avoids the propagation of the scalar mode present in the previous
derivations~\cite{Lu:2020iav,Kobayashi:2020wqy,Fernandes:2020nbq,Hennigar:2020lsl,Bonifacio:2020vbk}.
Symmetries that prohibit the scalar mode from propagating
have been discussed in Ref.~\cite{Tasinato:2020fni}.
See Refs.~\cite{Feng:2019dwu,Mukohyama:2019unx,Yao:2020tur}
for further developments in ``scalarless'' modified gravity
and Refs.~\cite{Aoki:2020oqc,DeFelice:2020prd,Sangtawee:2021mhz}
for the application to cosmology.

\subsection{Quasi-Einstein frame}

A field redefinition of a metric given by a disformal transformation~\cite{Bekenstein:1992pj}
retains the number of propagating DOFs as long as the transformation is invertible.
In terms of the ADM variables, it is expressed as
\begin{align}
N\to {\cal N}(t,N),\quad N^i\to N^i,\quad
\gamma_{ij}\to{\cal A}(t,N)\gamma_{ij}.
\end{align}
Let us consider a special case of a disformal transformation,
\begin{align}
N\to b_0(t)N+b_1(t),\quad N^i\to N^i,\quad \gamma_{ij}\to \gamma_{ij}.
\label{DT:special}
\end{align}
Under the transformation~\eqref{DT:special},
the form of the action~\eqref{action:Gao} is invariant,
with $\beta_A(t)$ and $\alpha_A(t)$ being transformed as
\begin{align}
&\beta_0\to\beta_0/b_0,\quad \beta_{1,2}\to (\beta_{1,2}+b_1)/b_0,
\notag \\ &
\alpha_{1,2}\to b_0\alpha_{1,2},\quad
\alpha_{3,4}\to \alpha_{3,4}+b_1\alpha_{1,2}.
\end{align}
Therefore, via the field redefinition~\eqref{DT:special} one can always set
$\beta_0=1$ and $\beta_2=0$.
Since the coefficient of $K_{ij}K^{ij}$ is now standard,
we dub this frame as the quasi-Einstein frame.
This motivates us to study the action of the form\footnote{As will be clear shortly,
we use units in which Newton's constant, $G_N$, is equal to $(8\pi\alpha_2)^{-1}$,
and we will be interested mainly in the case with $\alpha_2=1$.}
\begin{align}
S&=\frac{1}{2}\int\D t\D^3x\, N\sqrt{\gamma}\biggl[
K_{ij}K^{ij}-\frac{1}{3}
\left(\frac{2N}{\beta+N}+1\right)K^2
\notag \\ & \quad \quad
+\alpha_1+\alpha_2R+\frac{1}{N}\left(\alpha_3+\alpha_4R\right)
\biggr],\label{action:Gao2}
\end{align}
where $\beta:=\beta_1$,
as the case of particular interest.
General relativity with a cosmological constant is reproduced by setting
$\beta=\alpha_3=\alpha_4=0$, $\alpha_1=\,$ const, and $\alpha_2=1$.
Theories with $\beta\neq 0$ are not included in the (extended) cuscuton family,
and hence the impacts of $\beta$ have not been investigated so far in the literature.

Note that a subtlety arises when matter is present.
In this paper
we assume that matter is minimally coupled to gravity
in this quasi-Einstein frame.

Here we should point out the similarity and difference
between the action~\eqref{action:Gao2} and that of the infrared limit of
nonprojectable Ho\v{r}ava gravity~\cite{Horava:2009uw,Blas:2009qj,Blas:2010hb},
\begin{align}
  S&=\frac{1}{2}\int\D t\D^3x\, N\sqrt{\gamma}\bigl(
 K_{ij}K^{ij}-\lambda K^2
\notag \\ & \quad\quad
 +\alpha_1+\alpha_2 R +\eta  a_ia^i
  \bigr),\label{action:Horava}
\end{align}
where $a_i:=D_i\ln N$ and now all the coefficients are constants.
The same action is also obtained in the context of
Einstein-aether theory~\cite{Jacobson:2000xp} if the aether field is restricted to be
hypersurface orthogonal~\cite{Blas:2009ck,Jacobson:2010mx}.
While being mostly similar,
the essential difference can be perceived in the structure of
the coefficient of $K^2$.
In the case of $\eta=0$, nonprojectable Ho\v{r}ava gravity
has odd dimensionality of the phase space at each spacetime point and
hence is inconsistent~\cite{Li:2009bg,Henneaux:2009zb}.
By contrast, the action~\eqref{action:Gao2} is free of
such a trouble thanks to the particular structure of the coefficient of $K^2$.

\section{Solar-system tests and gravitational waves}\label{sec:GWs-Newton}

Let us study gravitational potentials produced by nonrelativistic matter
and the propagation of gravitational waves in the theory described
by the action~\eqref{action:Gao2}.
We assume that $\alpha_1, \alpha_3\sim H_0^2$, where $H_0$
is the Hubble scale, so that they are only relevant to
the large-scale cosmological dynamics.
We also assume that
$\alpha_2$ and $\alpha_4$ could vary only on cosmological time scales, because
otherwise nearly static gravitational potentials would not be possible.
We can then consider
scalar and tensor perturbations on a Minkowski background,
using the approximation
$\alpha_{1},\alpha_{3}\ll \partial_t^2,\Delta:=\delta_{ij}\partial_i\partial_j$.

Static scalar perturbations of the ADM variables are given by
\begin{align}
N=1+\Phi(\Vec{x}),\quad N_i=\partial_i\chi(\Vec{x}),
\quad \gamma_{ij}=[1-2\Psi(\Vec{x})]\delta_{ij}.
\end{align}
The quadratic action reads
\begin{align}
S&=\int\D t\D^3x\biggl[
-(\alpha_2+\alpha_4)\Psi\Delta\Psi+2\alpha_2\Phi\Delta\Psi
\notag \\ & \quad\quad
+\frac{\beta}{3(1+\beta)}(\Delta\chi)^2-\Phi\rho(\Vec{x})\biggr],
\label{newton-action}
\end{align}
where $\rho(\Vec{x})$ is the matter energy density and
this matter is assumed to be minimally coupled to gravity.
Note that in deriving Eq.~\eqref{newton-action}
we actually assumed that $\beta\Delta\gg H_0^2$.
The equations of motion are then solved to give
\begin{align}
\Delta\Phi=\left(1+\frac{\alpha_4}{\alpha_2}\right)
\frac{\rho}{2\alpha_2},
\quad
\Delta\Psi = \frac{\rho}{2\alpha_2},
\quad \Delta\chi=0.
\end{align}
This result implies that
\begin{align}
\alpha_2\simeq\mathrm{const}\quad\mathrm{and}\quad \alpha_4\simeq 0
\end{align}
are required in order to evade solar-system constraints
(light bending, the Shapiro time delay, and the time variation of Newton's
constant).\footnote{According to the current limits on
the time variation of Newton's constant, $|\dot\alpha_2/\alpha_2|$
must be much smaller than $H_0$~\cite{Will:2018bme}.}

Tensor perturbations are given by
\begin{align}
N=1,\quad N_i=0,\quad \gamma_{ij}=\delta_{ij}+h_{ij}(t,\Vec{x}),
\end{align}
for which the quadratic action is
\begin{align}
S=\frac{1}{8}\int\D t\D^3x \left[
(\partial_th_{ij})^2+(\alpha_2+\alpha_4)h_{ij}\Delta h_{ij}
\right].
\end{align}
The observation of GW170817 and GRB 170817A
put a tight bound on the speed of gravitational waves,
$c_{\mathrm{GW}}\simeq 1$~\cite{LIGOScientific:2017vwq,LIGOScientific:2017zic}.
It is therefore required that
\begin{align}
\alpha_2+\alpha_4\simeq 1.
\end{align}

Summarizing these results, we identify
a phenomenologically interesting class characterized by
\begin{align}
\alpha_2=1,\quad \alpha_4=0,
\end{align}
which evades solar-system and gravitational-wave constraints.
Note that in these noncosmological phenomena
we do not see any effects of nonvanishing $\beta$.

\section{Black hole solutions}\label{sec:black-hole}

In this section, we study black hole solutions
in the theory defined by the action~\eqref{action:Gao2}.
On the basis of the discussion in the previous section,
we are interested in the phenomenologically viable case with
$\alpha_2=1$ and $\alpha_4=0$. Black holes under consideration are
supposed to be much smaller than the size of the cosmological horizon.
Therefore, we are mostly interested in asymptotically flat
black holes obtained ignoring $\alpha_1$ and $\alpha_3$.
(These two functions are assumed to be characterized by the
cosmological horizon scale.)
The remaining unfixed function, $\beta$,
is assumed to be constant in this section, because
we seek for stationary black hole solutions.
Our focus is therefore on whether the effects of $\beta$ can be seen or not
in black hole solutions.
Nevertheless, having said that,
we allow $\alpha_1$ to be a nonvanishing constant and
$\alpha_2$ to be a constant different from 1 in the following discussion,
because relaxing these assumptions does not hinder us
from presenting an analytic solution below.
To sum up, in this section we have three constant parameters,
$\beta$, $\alpha_1$, and $\alpha_2$, though we are mostly interested
in the case with $\alpha_1=0$ and $\alpha_2=1$.

\subsection{Static and spherically symmetric solution}\label{subsec:bh}

The ADM variables for static and spherically symmetric solutions
are taken to be
\begin{align}
&N=N(r),\quad N_i\D x^i=B(r)F(r)\D r,
\notag \\ &
\gamma_{ij}\D x^i\D x^j=F^2(r)\D r^2+r^2\D\Omega^2,
\label{ADMv:black-hole}
\end{align}
where $\D\Omega^2=\D\theta^2+\sin^2\theta\D\varphi^2$.
Since the action is only invariant under spatial diffeomorphisms,
at this stage
we do not make a temporal coordinate transformation, $t\to T(t,r)$,
to remove $B$. Such a coordinate transformation would result in
an inhomogeneous configuration of the nondynamical scalar field.

Substituting the ADM variables~\eqref{ADMv:black-hole}
to the action and varying it with respect to $N$, $B$, and $F$,
we obtain
\begin{align}
&\alpha_1+\frac{2\alpha_2}{r^2}\left[\left(\frac{F^2-1}{F^2}\right)r\right]'
+\frac{2(rB^2)'}{r^2(\beta+N)^2F^2}
\notag \\ &
-\frac{2r^2}{3}
\frac{\beta(\beta+2N)}{N^2(\beta+N)^2F^2}\left[
\left(B/r\right)'
\right]^2=0,\label{BH:Ham}
\\
&\frac{r^2B}{(\beta+N)^2F^2}\left[(\beta+N)F\right]'
+\frac{\beta}{3}\left[
\frac{r^4(B/r)'}{N(\beta+N)F}
\right]'
=0,\label{BH:Mom}
\\
&\alpha_1\frac{N}{F}+\frac{2\alpha_2}{r^2}\frac{N}{F}
\left(\frac{F^2-1}{F^2}-\frac{2r N'}{NF^2}\right)
\notag \\ &
+\frac{2(rB^2)'}{r^2NF^3}
-\frac{2\beta}{3r^4N(\beta+N)F^3}\left[(r^2B)'\right]^2=0,\label{BH:Ev}
\end{align}
where a dash denotes differentiation with respect to $r$.
The system is
of second order for $B$ (provided that $\beta\neq 0$) and
of first order for $N$ and $F$.

Since it is difficult to obtain a general solution to
Eqs.~\eqref{BH:Ham}--\eqref{BH:Ev}, we start with finding a particular solution.
One can easily verify that Eqs.~\eqref{BH:Ham}--\eqref{BH:Ev} admit
the following solution:
\begin{align}
N=N_0\sqrt{f(r)},\quad F=\frac{1}{\sqrt{f(r)}},\quad B=\frac{N_0b_0}{r^2},
\label{soln:Sch}
\end{align}
where
\begin{align}
    f(r):=1+\frac{\alpha_1}{6\alpha_2}r^2-\frac{\mu_0}{r}+\frac{b_0^2}{\alpha_2r^4},
\end{align}
and $N_0$, $\mu_0$, and $b_0$ are integration constants.
Here, $N_0$ is physically less important because it may be set to 1
by rescaling the unit of $t$.

To see the spacetime structure of this solution more clearly,
we note that
the four-dimensional metric can be written as
\begin{align}
  \D s^2&=
-(N^2-B^2)\left(\D t-\frac{BF}{N^2-B^2}\D r\right)^2
\notag \\ &\quad  +\frac{N^2F^2}{N^2-B^2}\D r^2+r^2\D\Omega^2,
\end{align}
where
\begin{align}
N^2-B^2&=N_0^2\left(
1+\frac{\alpha_1}{6\alpha_2}r^2-\frac{\mu_0}{r}+\frac{1-\alpha_2}{\alpha_2}\frac{b_0^2}{r^4}\right),
\\
NF&=N_0.
\end{align}
Thus, introducing a new time coordinate defined by
$\D T=N_0\left\{\D t-[BF/(N^2-B^2)]\D r\right\}$, one can write the metric
in a diagonal form at the expense of a homogeneous configuration
of the nondynamical scalar field (see Appendix~\ref{app:covariantize}).
It is now obvious that in the phenomenologically interesting case of $\alpha_2=1$,
the solution~\eqref{soln:Sch} describes
Schwarzschild-(anti-)de Sitter spacetime. Importantly,
we see no $\beta$-dependence in this solution.
Note that inside the black hole horizon there is a location
at which $N=0$. If $\alpha_1<0$, such a location also exists outside
the cosmological horizon. These are called the universal horizon,
which has been studied in Einstein-aether theory and Ho\v{r}ava
gravity~\cite{Barausse:2011pu,Blas:2011ni}.

Aside from the cosmological constant term,
essentially the same expression for a black hole solution has been
obtained in Einstein-aether theory~\cite{Berglund:2012bu}.
This is not surprising because the solution~\eqref{soln:Sch}
satisfies the maximal slicing condition,
$K=-(r^2B)'/r^2NF=0$, while the crucial difference between the two theories
is found in the coefficient of $K^2$ in the action.
The maximal slicing condition is the very reason why
there is no $\beta$-dependence in the solution.
Note that a similar solution was found
in a different modified gravity theory with two
tensorial DOFs~\cite{DeFelice:2020onz}.

\subsection{Static perturbations}

We have thus obtained a particular solution~\eqref{soln:Sch},
but at this stage it is not clear whether or not there exist other black hole
solutions with the appropriate asymptotic behavior.
Since $B$ obeys the second-order differential equations,
one might rather expect that there should be another integration constant
in addition to $N_0$, $\mu_0$, and $b_0$.
To address this point, let us consider a slightly deformed solution,
\begin{align}
&N=N_0\sqrt{f+h_0(r)},\quad F=\frac{1}{\sqrt{f+h_1(r)}},
\notag \\ &
 B=N_0\left[\frac{b_0}{r^2}+h_2(r)\right],
\end{align}
where $h_1$, $h_2$, and $h_3$ are to be treated as small perturbations.

Equations~\eqref{BH:Ham} and~\eqref{BH:Ev} reduce, respectively, to
\begin{align}
&f\left[\frac{2b_0}{\alpha_2}\left(\frac{h_2}{r}\right)'
-(rh_1)'\right]+\frac{3b_0^2}{\alpha_2r^4}(h_0-h_1)=0,\label{eq:pert1}
\\
&f\left[\frac{2b_0}{\alpha_2}\left(\frac{h_2}{r}\right)'
-(rh_0)'\right]+\left(1+\frac{\alpha_1}{2\alpha_2}r^2\right)(h_0-h_1)=0.
\end{align}
Eliminating $h_2$ from these two equations, we obtain
\begin{align}
f(h_0-h_1)'-f'(h_0-h_1)=0,
\end{align}
which can be integrated to give
\begin{align}
h_0-h_1=C_1f,
\end{align}
where $C_1$ is an integration constant.
This integration constant can be absorbed into a redefinition of $N_0$.
Substituting this to Eq.~\eqref{eq:pert1},
we have
\begin{align}
h_0&=-\frac{\mu_1}{r}+C_1\left(f-\frac{b_0^2}{\alpha_2 r^4}\right)+\frac{2b_0}{\alpha_2}\frac{h_2}{r^2},
\\
h_1 &=-\frac{\mu_1}{r}-\frac{C_1b_0^2}{\alpha_2 r^4}+\frac{2b_0}{\alpha_2}\frac{h_2}{r^2},
\end{align}
where $\mu_1$ is an integration constant.
This integration constant simply corresponds to a constant shift of
the mass parameter $\mu_0$.
Finally, substituting all these results to Eq.~\eqref{BH:Mom}, we arrive at
\begin{align}
&\left(\beta+N_0\sqrt{f}\right)\left(h_2''+\frac{2}{r}h_2'-\frac{2}{r^2}h_2\right)
\notag \\ &
-N_0(\sqrt{f})'\left(h_2'+\frac{2}{r}h_2\right)=0,
\end{align}
where we assumed $\beta\neq 0$.
This equation can be integrated to give
\begin{align}
h_2=\frac{b_1}{r^2}+\frac{\epsilon_1}{r^2}\int^r r^2\left(\beta+N_0\sqrt{f}\right)\D r ,
\end{align}
where $b_1$ and $\epsilon_1$ are integration constants.
The former integration constant corresponds to a constant shift of $b_0$,
while the latter cannot be absorbed into a shift of the integration
constants of the solution~\eqref{soln:Sch}.
Let us now focus specifically on the case of a zero cosmological constant, $\alpha_1=0$,
where the solution is asymptotically flat at zeroth order.
In this case, we have
\begin{align}
h_2\sim \frac{\epsilon_1}{3}(\beta+N_0)r
\end{align}
for large $r$, which would invalidate the perturbative approximation
unless $\epsilon_1=0$.
Therefore, no asymptotically flat solution can be obtained
by a small deformation of the solution~\eqref{soln:Sch}.
Our result agrees with that of the decoupling limit analysis
of the khrononmetric theory~\cite{Lin:2017jvc}.

\subsection{Numerical solutions}

  \begin{figure}[tb]
    \begin{center}
        \includegraphics[keepaspectratio=true,height=58mm]{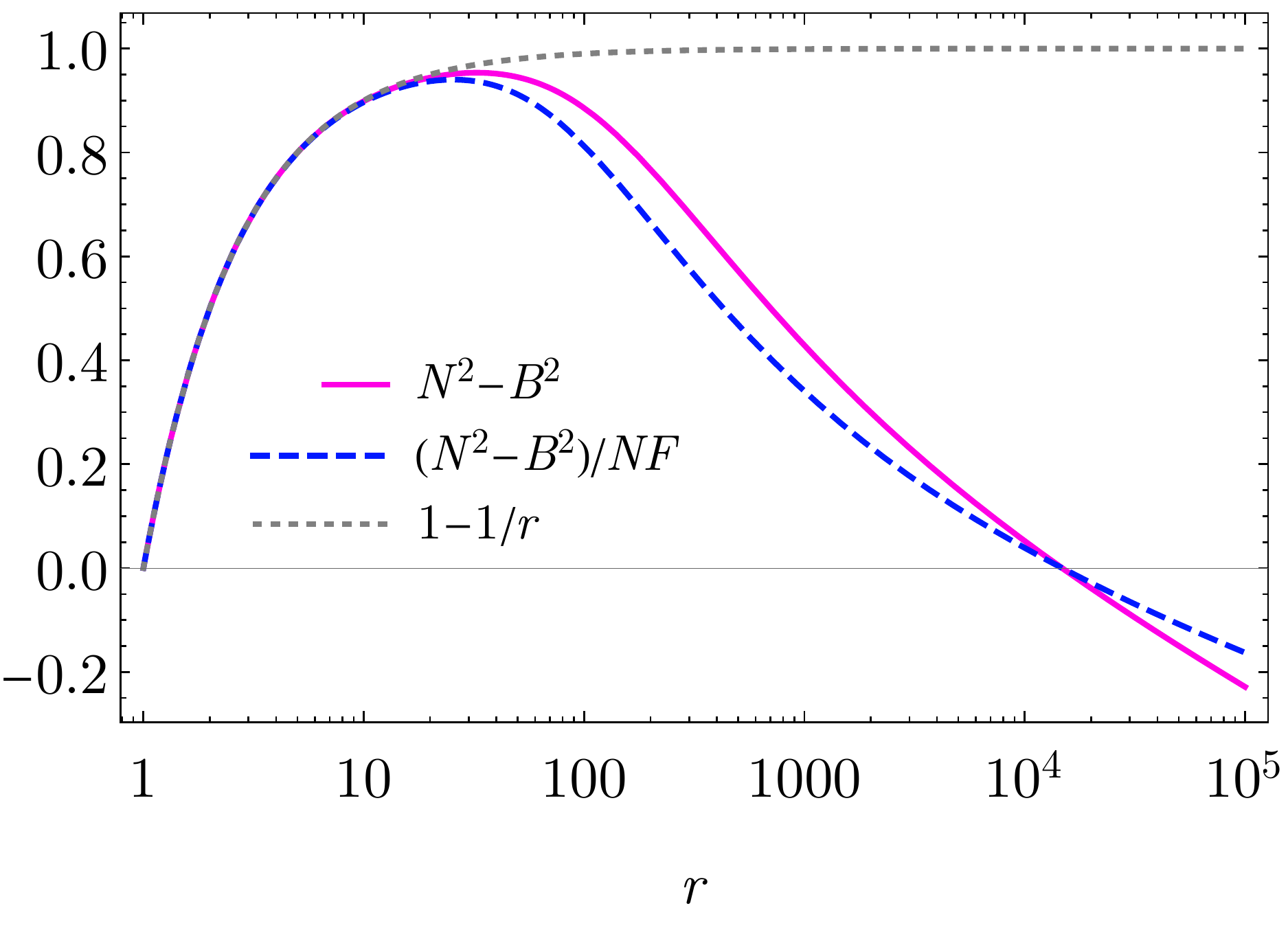}
    \end{center}
      \caption{Numerical solution for $r_h=1$, $N_h=0.1$, and $B_1=-1.9N_h$.
      The theory parameters are given by $\alpha_1=0$, $\alpha_2=1$, and $\beta=0.2$.
  	}
      \label{fig:flat1.pdf}
  \end{figure}

  \begin{figure}[tb]
    \begin{center}
        \includegraphics[keepaspectratio=true,height=58mm]{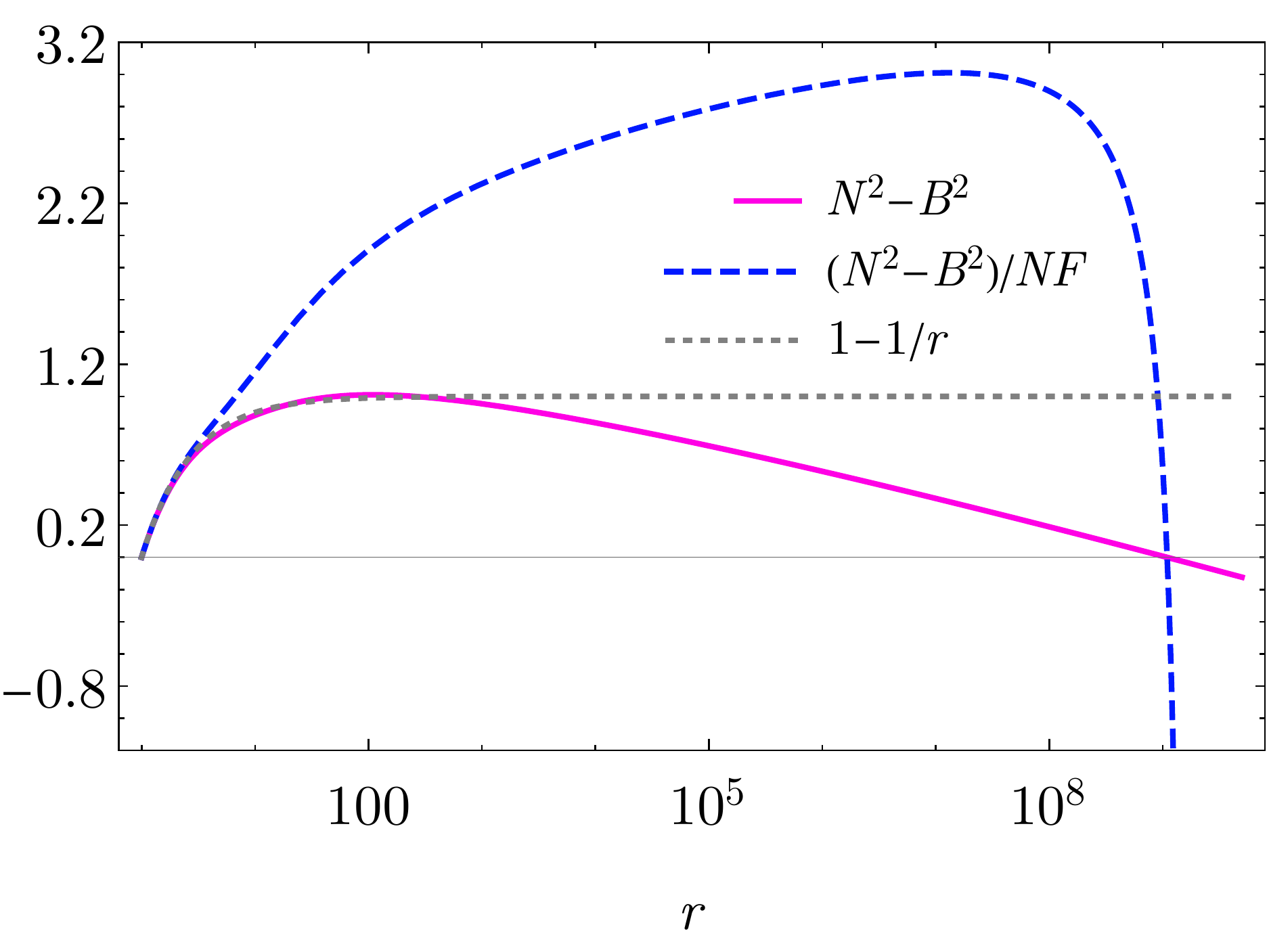}
    \end{center}
      \caption{Numerical solution for $r_h=1$, $N_h=0.1$, and $B_1=-3N_h$.
      The theory parameters are given by $\alpha_1=0$, $\alpha_2=1$, and $\beta=-0.3$.
  	}
      \label{fig:flat2.pdf}
  \end{figure}

We perform a numerical analysis that is complementary to
and can go beyond the perturbative study in the previous subsection.

In the vicinity of the horizon, $r=r_h$, the ADM variables can be expanded as
\begin{align}
N&=N_h+N_1(r-r_h)+\cdots,
\\
F&=\frac{1}{N_h}+F_1(r-r_h)+\cdots,
\\
B&=N_h+B_1(r-r_h)+\frac{B_2}{2}(r-r_h)^2+\cdots.
\end{align}
Here, one may set $F(r_h)=1/N_h$ without loss of generality
since doing so corresponds to fixing the scale of time.
For given $(r_h,N_h,B_1)$, one can numerically integrate Eqs.~\eqref{BH:Ham}--\eqref{BH:Ev}
from the horizon outwards to determine the profile of $N$, $F$, and $B$.
We thus try to find asymptotically flat solutions
that are not described as a perturbation of
the Schwarzschild spacetime foliated by maximal slices.

We consider again the case without the cosmological constant, $\alpha_1=0$.
For
\begin{align}
B_1=-\frac{2N_h}{r_h},\label{horizon:tune}
\end{align}
one has the asymptotically flat solution given by Eq.~\eqref{soln:Sch} (with $\alpha_1=0$).
As far as we have investigated numerically,
no asymptotically flat solutions have been found
if the condition~\eqref{horizon:tune} is not satisfied.\footnote{We have
not looked for asymptotically flat solutions in the case of
$\alpha_1\neq 0$.}
Typical examples obtained in the case of detuned conditions
are given in Figs.~\ref{fig:flat1.pdf} and~\ref{fig:flat2.pdf}.
The large $r$ behavior of these numerical solutions is also different
from that of asymptotically de Sitter solutions.

Our numerical analysis thus supports the conclusion that asymptotically flat,
spherically symmetric black holes in
the theory defined by the action~\eqref{action:Gao2}
with $\beta=\,$const, $\alpha_2=1$, and $\alpha_1=\alpha_3=\alpha_4=0$
are indistinguishable from those in GR.


\subsection{Slowly rotating solution}

Let us now discuss slowly rotating black hole solutions
(see Refs.~\cite{Barausse:2013nwa,Barausse:2015frm,Adam:2021vsk} for rotating
black holes in Einstein-aether theory and Ho\v{r}ava gravity).
To first order in the black hole spin, we write~\cite{Hartle:1967he,Hartle:1968si}
\begin{align}
  \delta N=0,\quad
\delta N_i\D x^i&=-r^2\sin^2\theta\,\omega(r,\theta)\D\varphi,
\quad \delta\gamma_{ij}=0.
\end{align}
The $(r\varphi)$ component of the evolution equations
reads
\begin{align}
  \partial_\theta\left(\sin^3\theta\partial_\theta\omega\right)
  =0,\label{rot:eq1}
\end{align}
which, together with the regularity at the pole, yields $\omega=\omega(r)$.
Then, the $\varphi$ component of the momentum constraints reduces to
\begin{align}
    \left(\frac{r^4\omega'}{NF}\right)'=0.
  \label{rot:eq2}
\end{align}
This can be integrated to give
\begin{align}
\omega(r) = -\frac{3J}{4\pi}\int^r\frac{NF}{r^4}\D r + \omega_0,
\end{align}
where $J$ and $\omega_0$ are integration constants.
Upon using $NF=1$ we arrive at $\omega = J/4\pi r^3+\omega_0$.
(For simplicity, here we have set $N_0=1$ without loss of generality.)
Noting that the time coordinate $t$ that we are using is nonstandard,
we perform a coordinate transformation
\begin{align}
\varphi\to \varphi+\int^r\frac{BF}{N^2-B^2}(\omega-\omega_0)\D r+\omega_0 t.
\end{align}
The four-dimensional metric then becomes
\begin{align}
\D s^2&=-(N^2-B^2)\D T^2+\frac{\D r^2}{N^2-B^2}+r^2\D\Omega^2
\notag \\ &\quad -\frac{J\sin^2\theta}{2\pi r}\D T\D\varphi,
\end{align}
which is nothing but the Kerr metric in GR
under the slow-rotation approximation.
Thus, we conclude that no effect of $\beta$ can be seen
in slowly rotating black holes.

\section{Cosmology}\label{sec:cosmology}

Our next step is the analysis of cosmology.
In this section, we will mainly consider the phenomenologically most
interesting case with the time-dependent functions
\begin{align}
    &\beta_0=1,\quad\beta_1=\beta(t),\quad \beta_2=0,
    \notag \\ 
    &\alpha_1=\alpha_1(t),\quad\alpha_2=1,\quad  \alpha_3=\alpha_3(t),\quad 
    \alpha_4=0.
\end{align}
However, when deriving a quadratic action for
cosmological perturbations, we will work in the general action~\eqref{action:Gao}
where no assumptions are made on these functions,
because the general form of the quadratic action will be used
for another purpose in the next section.

To investigate cosmology, we add a (irrotational, barotropic)
perfect fluid which is minimally coupled
to gravity. Such a fluid component can be mimicked by a scalar field
whose action is given by
\begin{align}
    S_{\mathrm{mat}}=\int\D^4x\sqrt{-g}P(Y),\quad Y:=-\frac{1}{2}(\partial\varphi)^2.
    \label{action:fluid}
\end{align}
The matter energy-momentum tensor reads
\begin{align}
    T_{\mu\nu}=(\rho+P)u_\mu u_\nu +Pg_{\mu\nu},
\end{align}
where the energy density and the four-velocity are given respectively by
\begin{align}
    \rho=2YP_{,Y}-P,\quad u_\mu=-\frac{\nabla_\mu\varphi}{\sqrt{2Y}}.
\end{align}
The equation of motion for $\varphi$ is equivalent to
the conservation law, $\nabla_\mu T_\nu^{\mu}=0$.
Here, $\nabla_\mu$ stands for the usual four-dimensional covariant derivative.
The sound speed of the fluid is given by
\begin{align}
    c_s^2=\frac{P_{,Y}}{P_{,Y}+2YP_{,YY}},
\end{align}
where the right hand side is evaluated at the homogeneous background.
For $P\propto Y^{(1+w)/2w}$ with $w=\,$const, we have
$c_s^2=P/\rho=w$.

\subsection{Homogeneous and isotropic background}\label{subsec:cosmo}

The total action is now composed of the gravity sector~\eqref{action:Gao2}
and the action for a cosmological fluid~\eqref{action:fluid}.

The ADM variables for a homogeneous and isotropic universe are given by
\begin{align}
    N=\bar N(t),\quad N_i=0,\quad \gamma_{ij}=a^2(t)\delta_{ij}.
\end{align}
Since time reparametrization symmetry is spontaneously broken,
one cannot put $\bar N=1$ in general, but,
as we will show below, the lapse function is determined
from the equations of motion.
From the Hamiltonian constraint and the evolution equations, we obtain
\begin{align}
    &\frac{3H^2}{(\beta/\bar N+1)^2}+\frac{\alpha_1}{2}=\rho,\label{cosmo:H}
    \\
    &-\frac{3H^2}{\beta/\bar N+1}-\frac{2}{\bar N}
    \frac{\D}{\D t}\left(\frac{H}{\beta/\bar N+1}\right)
    -\frac{1}{2}\left(\alpha_1+\frac{\alpha_3}{\bar N}\right)=P,\label{cosmo:E}
\end{align}
where
$H:=\bar N^{-1}\D \ln a/\D t$
is the Hubble parameter.
(As far as the spatially flat model is concerned, the coefficients of $R$,
i.e., $\alpha_2$ and $\alpha_4$, do not appear anyway in the equations for
a homogeneous universe.)
The conservation law reads
\begin{align}
    \bar N^{-1}\dot \rho+3H(\rho+P)=0,\label{cosmo:cons}
\end{align}
where a dot stands for differentiation with respect to $t$.
Given the equation of state, this equation can be integrated to
give $\rho$ and $P$ as a function of $a$: $\rho=\rho(a)$, $P=P(a)$.
Note that Eq.~\eqref{cosmo:cons}
is not an automatic consequence of Eqs.~\eqref{cosmo:H}
and~\eqref{cosmo:E}. Rather, substitution of Eqs.~\eqref{cosmo:H}
and~\eqref{cosmo:E} into Eq.~\eqref{cosmo:cons} yields
\begin{align}
    &\dot\alpha_1-3H\alpha_3
    -\frac{6\beta H}{\beta/\bar N+1}
    \left[
    \frac{3H^2}{\beta/\bar N+1}+\frac{2}{\bar N}
    \frac{\D}{\D t}\left(\frac{H}{\beta/\bar N+1}\right)
    \right]
    \notag \\ &
    =0.
\end{align}
Using Eqs.~\eqref{cosmo:H} and~\eqref{cosmo:E} again, one can
write this equation as the constraint
among the time-dependent functions,
\begin{align}
    \dot\alpha_1-\sqrt{3}\left(\rho-\frac{\alpha_1}{2}\right)^{1/2}
    \left[\alpha_3-(\alpha_1+2P)\beta\right]=0.
    \label{cosmo:cusuc}
\end{align}
In the covariant formulation presented in Appendix~\ref{app:covariantize},
this follows from the equation of motion for a St\"{u}ckelberg scalar field
(see Eq.~\eqref{app:eom-stuck}).
In the ADM formulation of usual scalar-tensor theories with two tensorial and one scalar DOFs,
the equation corresponding to Eq.~\eqref{cosmo:cusuc}
reduces to a first-order differential equation for the lapse function, which can
be used to determine $\bar N=\bar N(t)$.
In contrast, in the present case Eq.~\eqref{cosmo:cusuc} does not contain $\bar N$,
which is the crucial point in ``scalarless'' theories.
Given that $\rho$ and $P$ are now expressed in terms of $a$,
Eq.~\eqref{cosmo:cusuc} instead allows us to write $a$ in terms of $t$.
Finally, substituting $a=a(t)$ into Eq.~\eqref{cosmo:H},
one can determine the lapse function $\bar N$ as a function of $t$.

Let us now present an example.
Suppose that the universe is filled with a fluid
with a constant equation of state parameter, $w=P/\rho$, and 
the time-dependent functions are given by
\begin{align}
    \beta&=\mathrm{const},\quad \alpha_2=1,\quad \alpha_4=0,
    \\
    \alpha_1&=6h_0^2\left[
    \frac{1}{\xi}-\frac{1}{(1+\beta)^2}\right]
    \coth^2\left[\frac{3}{2}(1+w)h_0t\right]
    -\frac{6h_0^2}{\xi},
    \\
    \alpha_3&=6h_0^2
    \left[\frac{1+w(1+\beta)}{(1+\beta)^2}
    -\frac{1+w}{\xi}\right]
     \mathrm{csch}^2\left[\frac{3}{2}(1+w)h_0t\right]
     \notag \\ & \quad 
    -\frac{6\beta h_0^2}{(1+\beta)^2},
\end{align}
where $\xi$ and $h_0$ are constant parameters.
This example admits the following solution:
\begin{align}
    \bar N=1,
    \quad 
    a^{3(1+w)}\propto 
    \sinh^2\left[\frac{3}{2}(1+w)h_0t\right].
\end{align}
Noting that $\alpha_1$ can be written as
\begin{align}
    \alpha_1=6\left[
    \frac{1}{\xi}-\frac{1}{(1+\beta)^2}\right]
    H^2-6\frac{h_0^2}{\xi},
\end{align}
we see that Eqs.~\eqref{cosmo:H} and~\eqref{cosmo:E} read
\begin{align}
    3 H^2&=\xi \rho+3 h_0^2,
    \\
    -2 \dot H &=\xi(\rho+P).
\end{align}
This shows that the cosmological gravitational constant differs from
Newton's constant $G_N$ by the factor $\xi$:
\begin{align}
    \frac{G_{\mathrm{cos}}}{G_N}=\xi.
\end{align}
(Recall that we are working in units where $8\pi G_N=1\,(=1/\alpha_2)$.)
Aside from this modification, the background evolution obeys the
standard Friedmann equations in the presence of a cosmological constant.
A similar situation occurs in Einstein-aether theory
and Ho\v{r}ava gravity, where
a mild bound on the difference between $G_{\mathrm{cos}}$ and $G_{N}$ has been
obtained from the measurement of the primordial abundance of
He$^4$ as $|\xi-1|\lesssim 0.1$~\cite{Carroll:2004ai,Blas:2009qj}.
In our example, $\xi$ is a free parameter so that one may set $\xi=1$,
leading to the background evolution that is completely indistinguishable
from that in GR.

In the above example, $\alpha_1$ and $\alpha_3$ track the
cosmic evolution as $\alpha_1,\alpha_3\sim H^2$.
We present another example in which $\alpha_1$ and $\alpha_3$
instead remain as small as $H_0^2$ and simply behave as a
dark energy component with a constant equation
of state parameter $w_{\mathrm{DE}}$. In the case where
the matter component is a pressureless fluid,
the model is given by
\begin{align}
    \beta&=\mathrm{const},\quad \alpha_2=1,\quad \alpha_4=0,
    \\
    \alpha_1&=-2\rho_{\mathrm{DE}}(t),
    \\
    \alpha_3&=2[1+w_{\mathrm{DE}}(1+\beta)]\rho_{\mathrm{DE}}(t),
    \\
    \rho_{\mathrm{DE}}&\propto A^{-3(1+w_{\mathrm{DE}})},
\end{align}
where $A$ and $t$ are related via
\begin{align}
     t&=\frac{2\sqrt{1+r_0}}{3H_0}A^{3/2}
    \notag \\ & \quad \times 
    {}_2F_1
    (1/2,-1/2w_{\mathrm{DE}},1-1/2w_{\mathrm{DE}},-r_0A^{-3w_{\mathrm{DE}}}).
\end{align}
Here, ${}_2F_1$ is the hypergeometric function
and $r_0:=\rho_{\mathrm{DE}}/\rho|_{a=1}$ is the present value of
the ratio between the dark energy density and the matter energy density.
One can check that the solution is given by $\bar N=1$ and $a=A(t)$.
This example yields the background evolution subject to
\begin{align}
    3H^2&=(1+\beta)^2(\rho+\rho_{\mathrm{DE}}),
    \\
    -2\dot H&=(1+\beta)^2\left[\rho+(1+w_{\mathrm{DE}})\rho_{\mathrm{DE}}
    \right],
\end{align}
i.e., it mimics the cosmology with
$G_{\mathrm{cos}}=(1+\beta)^2G_N$ and the dark energy component.

\subsection{Cosmological perturbations}

We go on to the analysis of scalar perturbations around a cosmological background,
with a particular focus on the case of a pressureless fluid.
The perturbed ADM variables are given by
\begin{align}
    N=\bar N(1+\delta n),\quad N_i=\bar N\partial_i\chi,
    \quad \gamma_{ij}=a^2e^{-2\psi}\delta_{ij},
\end{align}
where we used the spatial gauge degrees of freedom
to write $\gamma_{ij}$ in the above form.

For the moment we work in the general action~\eqref{action:Gao}
to derive the quadratic action for the scalar perturbations,
and later we will focus on the specific case of our interest.
The action~\eqref{action:Gao} can be expanded to quadratic order in perturbations as
\begin{align}
    S^{(2)}&=\int\D t\D^3x\bar N a^3\biggl[-\frac{3\beta_0}{\bar \beta_1}
    \left(\psi_n+\frac{H\delta n}{\bar\beta_1}\right)^2
    \notag \\ &\quad 
    -\frac{2\beta_0}{\bar\beta_1}\left(\psi_n
    +\frac{H\delta n}{\bar \beta_1 }\right)\frac{\Delta\chi}{a^2}
    +\frac{\beta_0(\bar\beta_1-\bar\beta_2)}{3\bar\beta_1\bar\beta_2}
    \left(\frac{\Delta\chi}{a^2}\right)^2
    \notag \\ & \quad 
    -\left(\alpha_2+\frac{\alpha_4}{\bar N}\right)\frac{\psi\Delta\psi}{a^2}
    +2 \alpha_2\delta n\frac{\Delta\psi}{a^2}\biggr],
    \label{pertaction:grav}
\end{align}
where the notations $\psi_n:=\bar N^{-1}\partial_t\psi$
and $\bar\beta_{1,2}:=\beta_{1,2}/\bar N+1$
are used.
The matter action~\eqref{action:fluid} can also be expanded in terms of perturbations as
\begin{align}
    S_{\mathrm{mat}}^{(2)}&=\int\D t\D^3x \bar N
    a^3\left(\frac{\rho+P}{c_s^2}\right)
    \biggl[c_s^2\frac{\delta\varphi}{\varphi_n}\frac{\Delta\chi}{a^2}+\frac{\delta n^2}{2}
    \notag \\ & \quad 
    -\frac{\delta\varphi_n}{\varphi_n}\delta n-3c_s^2\frac{\delta\varphi_n}{\varphi_n}\psi
    +\frac{1}{2}\left(\frac{\delta\varphi_n}{\varphi_n}\right)^2
    \notag \\ & \quad 
    +\frac{c_s^2}{2}\frac{\delta\varphi\Delta\delta\varphi}{\varphi_n^2a^2}
    \biggr],\label{pertaction:mat}
\end{align}
where $\delta\varphi$ is the perturbation of $\varphi$.
Our total action is given by
$S_{\mathrm{tot}}^{(2)}=S^{(2)}+S_{\mathrm{mat}}^{(2)}$.\footnote{We omit
$\delta n\psi$ and $\psi^2$ from Eqs.~\eqref{pertaction:grav}
and~\eqref{pertaction:mat} because upon combining $S^{(2)}$ and 
$S^{(2)}_{\mathrm{mat}}$ they can be eliminated in the end by the
use of the background equations.}
This is the complete expression for the quadratic action
derived from the general action~\eqref{action:Gao} in the
presence of a perfect fluid (or a shift-symmetric k-essence field).

Now we focus on
the case of our interest: $\beta_0=\alpha_2=1$ and $\beta_2=\alpha_4=0$
with the notation $\beta_1=\beta$.
For simplicity, we consider the background with $\bar N=1$.
We follow the analysis of density perturbations in the extended
cuscuton theory~\cite{Iyonaga:2020bmm} to derive the reduced action
written solely in terms of a single variable $\delta$ defined by
\begin{align}
    \delta=\frac{\rho+P}{\rho c_s^2}
    \left(\frac{\dot{\delta\varphi}}{\dot\varphi}-\delta n\right)
    -\frac{3(\rho+P)}{\rho}\psi,\label{def:delta}
\end{align}
where note that the density perturbation is given by
$\delta\rho=(\rho+P)/c_s^2\cdot (\dot{\delta\varphi}/\dot\varphi-\delta n)$.
Later at an appropriate point
we will carefully take the limit $c_s^2,P\to 0$ while keeping $\rho$ finite.
We start with introducing $\delta$ as an auxiliary field
and write $S_{\mathrm{tot}}^{(2)}$ equivalently as
\begin{align}
    S_{\mathrm{tot}}^{(2)}&=S^{(2)}+S_{\mathrm{mat}}^{(2)}
    -\int\D t\D^3x a^3
    \left(\frac{\rho+P}{2c_s^2}\right)
    \notag \\ &\quad \times 
    \biggl[
    \frac{\dot{\delta\varphi}}{\dot\varphi}-\delta n 
    -c_s^2\left(\frac{\rho \delta}{\rho+P}+3\psi\right)
    \biggr]^2.\label{qaction:aux}
\end{align}
The additional third term is introduced so that it
removes $\dot{\delta\varphi}{}^2$ in $S_{\mathrm{tot}}^{(2)}$.
Namely, one ends up with the action that depends linearly on $\dot{\delta\varphi}$.
The equation of motion for $\delta$ yields Eq.~\eqref{def:delta}.
Substituting this to Eq.~\eqref{qaction:aux}, the third term vanishes and the
action reduces back to the original one, $S^{(2)}+S_{\mathrm{mat}}^{(2)}$.
Thus, the two representations are indeed equivalent.

Since $S_{\mathrm{tot}}^{(2)}$ is now linear in $\dot{\delta\varphi}$,
one can use the equation of motion to remove $\delta\varphi$ from the action.
At this stage one may substitute $P\propto Y^{(1+w)/2w}$ and
express the action in terms of $\rho$ and $w$,
which allows us for a $w\to 0$ limit without any divergences.
Then, $\delta n$ and $\chi$ can also be removed from the action
by the use of their equations of motion.
The resulting action is of the form
\begin{align}
    S_{\mathrm{tot}}^{(2)}&=\int \D t\D^3x a^3\bigl[
    \dot\delta{\cal A}(t,\Delta )\dot\delta+\delta {\cal O}_1(t,\Delta )\delta
    \notag \\ &\quad 
    +2\psi {\cal O}_2(t,\Delta)\delta+\psi {\cal O}_3(t,\Delta)\psi
    \bigr],
\end{align}
where 
\begin{align}
    {\cal A}=\frac{\rho}{3\rho-2\Delta/a^2}.
\end{align}
The explicit expressions for ${\cal O}_1$, ${\cal O}_2$, and ${\cal O}_3$ are messy
and not illuminating. Finally, one can eliminate
$\psi$ by using its equation of motion and arrives at the reduced action
expressed in terms of $\delta$ alone:
\begin{align}
    S_{\mathrm{tot}}^{(2)}&=\int \D t\D^3x a^3\left[ 
    \dot\delta{\cal A}(t,\Delta)\dot\delta+\delta {\cal B}(t,\Delta)\delta
    \right].
\end{align}

On the basis of this action we move to study the behavior of
the perturbations in the short and long wavelength limits.
In both limits,
we will see that the effect of modification of gravity comes into play
only through the underlying background model.

\subsubsection{Short wavelength limit}

In the short wavelength limit, $\Delta/a^2\gg \rho$, we have
\begin{align}
    {\cal A}\simeq -\frac{1}{2}a^2\rho\Delta^{-1},
    \quad 
    {\cal B}\simeq -\frac{1}{4}a^2\rho^{2}\Delta^{-1}.
\end{align}
The time-dependent function in the action $\beta$ has dropped out.
Since $\rho\propto a^{-3}$, the equation of motion for $\delta$ is found to be
\begin{align}
    \ddot \delta+2H\dot\delta = \frac{1}{2}\rho\delta.
\end{align}
Recalling that $4\pi G_N=1/(2\alpha_2)=1/2$,
this is identical to the evolution equation for matter density perturbations
in GR. Note, however, that the evolution of $\delta$ depends
on the underlying background model, which could be different from
that of the $\Lambda$CDM model. For example, if $G_{\mathrm{cos}}/G_N\neq 1$
as in the example in the previous subsection, the actual evolution of
the density perturbations would be modified.

The equations of motion for $\delta n$, $\chi$, and $\psi$
allow us to write these variables in terms of $\delta$.
In the short wavelength limit, we obtain
\begin{align}
    \delta n&\simeq \frac{1}{2}a^2\rho\Delta^{-1}\delta,
    \\
    \chi&\simeq 
    -\frac{3a^4\rho}{2(1+\beta)}\left[H\Delta^{-2}\delta+(1+\beta)\Delta^{-2}\dot\delta\right],
    \\
    \psi&\simeq \frac{1}{2}a^2\rho\Delta^{-1}\delta
    .
\end{align}
Therefore, the metric potentials that are used in
the familiar Newtonian gauge analysis,
\begin{align}
    \Phi&:=\delta n +\dot\chi,
    \\
    \Psi &:=\psi -H\chi,
\end{align}
obey the standard relation,
\begin{align}
    \Delta\Phi=\Delta\Psi=\frac{a^2}{2}\rho\delta,
\end{align}
with $\delta \simeq \delta\rho/\rho$.
Note that if one ignores the cosmic expansion, then
this reproduces the result obtained in Sec.~\ref{sec:GWs-Newton}.

\subsubsection{Long wavelength limit}

In the long wavelength limit, $\Delta/a^2\ll \rho$, we have 
\begin{align}
    {\cal A}\simeq \frac{1}{3},\quad {\cal B}\simeq -\frac{1}{9}\frac{\Delta}{a^2}.
\end{align}
Again, $\beta$-dependence has dropped out.
The long wavelength solution to the equation of motion for $\delta$
is given by
\begin{align}
    \delta\simeq \delta_0(\Vec{x})-\frac{\Delta\delta_0(\Vec{x})}{3}\int^t\frac{\D t'}{a^3(t')}
    \int^{t'}\D t''a(t''),
\end{align}
where $\delta_0$ is independent of time and the decaying mode has been discarded.
From the equations of motion for the other variables
in the long wavelength limit, we obtain
\begin{align}
    \delta n&={\cal O}(\Delta\delta/a^2H^2),
    \\
    \chi&\simeq a^2\Delta^{-1}\dot\delta,
    \\
    \psi&\simeq -\frac{\delta}{3},
\end{align}
leading to
\begin{align}
    \Phi=\Psi=-\frac{\delta_0(\Vec{x})}{3}\left[ 
    1-\frac{H}{a}\int^t a(t')\D t'
    \right].\label{long-solution}
\end{align}
The solution expressed in this way is the same as the long wavelength solution
for $\Phi$ and $\Psi$ in the presence of
a pressureless fluid and a cosmological constant in GR.
In particular, in the matter-dominated era, $a\propto t^{2/3}$, we have
the time-independent metric potentials, $\Phi=\Psi=-\delta_0/5$.
Note, however, that Eq.~\eqref{long-solution} has been derived
without assuming any particular form of $a(t)$ for an 
accelerated phase of the cosmic expansion.
In our modified theory of gravity, the background evolution
(of the accelerated phase) could be
different from that of the $\Lambda$CDM model depending on the
time-dependent functions in the action, so that the actual time-dependence
of $\Phi$ and $\Psi$ could be different from that in the $\Lambda$CDM model
away from the matter-dominated era.

\section{When matter matters}\label{sec:matter}

Finally, we discuss the problem of the coupling to generalized matter fields.
Let us consider the cubic galileon~\cite{Nicolis:2008in}
whose action is given by
\begin{align}
    S_{\mathrm{gal}}=\int\D^4x\sqrt{-g}\left[
    -\frac{1}{2}(\partial\varphi)^2-\frac{c}{2}(\partial\varphi)^2\Box\varphi\right],
    \label{action:gal}
\end{align}
where $c$ is a nonzero constant. In the following we will show that
the extra dynamical scalar DOF would reappear when one adds to
the action~\eqref{action:Gao} the galileon as a matter field.
This fact can be demonstrated by studying a quadratic action
for scalar perturbations around a cosmological background.

Expanding the action~\eqref{action:gal} to second order in perturbations,
we obtain
\begin{widetext}
\begin{align}
    S_{\mathrm{gal}}^{(2)}&=\int\D t\D^3x\bar N a^3\biggl[
    \left(\varphi_n-3c H\varphi_n^2\right)\delta\varphi\frac{\Delta\chi}{a^2}
    +c\varphi_n^2\delta\varphi_n\frac{\Delta\chi}{a^2}-c\varphi_n^3\delta n\frac{\Delta\chi}{a^2}
    -c\varphi_n^2\delta n\frac{\Delta\delta\varphi}{a^2}
    +\frac{\varphi_n^2}{2}\left(1-12c H\varphi_n\right)\delta n^2
    \notag \\ & \quad 
    -\varphi_n\left(1-9c H\varphi_n\right)\delta n\delta\varphi_n
    -3c\varphi_n^3\delta n\psi_n
    -3\left(\varphi_n-3c H\varphi_n^2\right)\psi\delta\varphi_n
    +3c\varphi_n^2\psi_n\delta\varphi_n
    +\frac{1}{2}\left(1-6c H\varphi_n\right)\delta\varphi_n^2
    \notag \\ & \quad
    +\frac{1}{2a^2}\left(1-4c H\varphi_n-2c\varphi_{nn}\right)\delta\varphi 
    \Delta\delta\varphi\biggr],
\end{align}
\end{widetext}
where $\delta\varphi$ is the fluctuation of the galileon field.
We add this to the action~\eqref{pertaction:grav}:
$S_{\mathrm{tot}}^{(2)}=S^{(2)}+S^{(2)}_{\mathrm{gal}}$.
To highlight what causes the problem, we do not make any further simplification.

Using the equations of motion for $\delta n$ and $\chi$, one can eliminate
them from the action. The resultant quadratic action is of the form
\begin{align}
    S^{(2)}_{\mathrm{tot}}=\int\D t\D^3x \bar N a^3\left({\cal K}_{IJ}q^I_nq^J_n+\cdots\right),
\end{align}
where $q^I=\{\psi,\delta\varphi\}$ and we wrote only the terms
quadratic in time derivatives. A straightforward calculation yields
\begin{align}
    \mathrm{det}\,
    {\cal K}&=-\frac{27[c\beta_0(\beta_1/\bar N) H\varphi_n^2]^2}{\Xi},
\end{align}
with
\begin{align}
    \Xi&:=12\beta_0^2H^2-2\beta_0\bar\beta_1^2(\bar\beta_1-\bar\beta_2)\varphi_n^2
    \notag \\ &\quad 
    +12c\beta_0\bar\beta_1[2\bar\beta_1(\bar\beta_1-\bar\beta_2)
    +\bar\beta_2]H\varphi_n^3
    \notag \\ & \quad 
    +3c^2\bar\beta_1^3\bar\beta_2\varphi_n^6.
\end{align}
This result shows that there are \textit{two} scalar DOFs
in the system unless $c\beta_1=0$; the scalar DOF reappears
in the presence of a galileon field
if it is minimally coupled to gravity in the frame where $\beta_1\neq 0$.

What we have seen here is analogous to the issue pointed out
in the context of degenerate higher-order
scalar-tensor theories~\cite{Deffayet:2020ypa,Garcia-Saenz:2021acj}.
In the case of the degenerate higher-order scalar-tensor theories,
the constraint associated with the degeneracy is lost
if a matter field is coupled to the Christoffel symbol.
In the present case, the constraint that removes the scalar DOF in the gravity sector
is lost due to the coupling of the cubic galileon to the Christoffel symbol.

\section{Conclusions and outlook}\label{sec:conclusions}

In this paper, we have studied aspects of spatially covariant theories of gravity
with two tensorial degrees of freedom (DOFs), namely, modified gravity
having the same number of DOFs as general relativity (GR).
We have mainly focused on a subset of the general theories developed in Ref.~\cite{Gao:2019twq}
that is characterized by three time-dependent functions. Two of the three functions
are related to the cuscuton terms~\cite{Afshordi:2006ad}
expressed in the unitary gauge, while the remaining one (denoted as $\beta$) is not
included even in the extended cuscuton theory~\cite{Iyonaga:2018vnu}.
More specifically, we have worked with the action~\eqref{action:Gao2}
with $\alpha_2=1$ and $\alpha_4=0$
as a particularly interesting subclass of the general action~\eqref{action:Gao},
with the purpose of exploring how the theory
can be distinguished observationally or phenomenologically from GR.

First, we have seen that the theory of modified gravity under consideration can
evade solar-system tests. We have also seen that
the speed of gravitational waves is equal to the speed of light.
Therefore, there is no obvious contradiction with observations and experiments
at this point.

Next, we have studied black hole solutions ignoring the cuscuton terms
that are supposed to be relevant only on cosmological scales.
Since the only modification in the action appears in the coefficient of
$K^2$, where $K$ is the trace of the extrinsic curvature of
constant time hypersurfaces, the theory admits GR solutions foliated by maximal slices
($K=0$). Therefore, we have the Schwarzschild solution.
The situation here is essentially the same as that in Einstein-aether theory
and the infrared limit of Ho\v{r}ava gravity.
We have considered small deformations of the Schwarzschild solution,
showing that no other asymptotically flat solutions can be obtained
by a perturbative treatment. Our numerical analysis beyond the perturbative treatment
supports the conclusion that the Schwarzschild solution is the only static,
spherically symmetric, and asymptotically flat vacuum solution.
We have also considered a slowly rotating black hole and obtained
the Kerr solution in the slow-rotation approximation.

We have then investigated the cosmological dynamics of the homogeneous
and isotropic background and scalar perturbations.
Since we have a large degree of freedom to choose the time-dependent
functions in the action including the cuscuton part, it is easy to
realize the background evolution that is very close to or even identical to
that of the $\Lambda$CDM model in GR.
As far as the long and short wavelength limits are concerned, we have found that
the perturbation dynamics is modified only through the modification
of the background evolution.

Combining these results, we conclude that it is quite difficult to
distinguish modified gravity with the action~\eqref{action:Gao2}
with $\alpha_2=1$ and $\alpha_4=0$
observationally or phenomenologically from GR.

Finally, we have noted, as a side remark, that in the presence of
a galileon field in the matter sector, the dynamical scalar degree of freedom
eliminated from the gravitational sector could reappear
in a similar way to the case analyzed in
Ref.~\cite{Deffayet:2020ypa,Garcia-Saenz:2021acj}.
The result of our analysis implies that a galileon field must be coupled
nonminimally to gravity in the frame where
the action takes the form of Eq.~\eqref{action:Gao2}.

Let us comment on several issues that are left for future work.
\begin{itemize}
    \item Newtonian gravity is reproduced in the weak field regime, and
    black hole solutions in GR are obtained in the strong field regime in vacuo.
    It is then natural to move on to the study of the structure of relativistic stars.
    \item It would be interesting to study black hole perturbations and
    quasinormal modes, which could help us to distinguish modified gravity
    with two tensorial DOFs from GR.
    \item In the present paper, we have studied cosmological perturbations
    only in the long and short wavelength limits. On intermediate scales
    the perturbation evolution could differ from that in GR even if
    the background evolution coincides with that of the $\Lambda$CDM model.
    A more detailed investigation is necessary.
    \item An application to inflationary cosmology would also be interesting.
\end{itemize}
These issues will be addressed in our forthcoming papers.

\acknowledgments
The work of AI was supported by the JSPS Research Fellowships for Young Scientists No.~20J11285.
The work of TK was supported by MEXT KAKENHI Grant Nos.~JP20H04745 and~JP20K03936.

\appendix

\section{Equations of motion}\label{app:eom}

The equations of motion in vacuum
derived from the action~\eqref{action:Gao2} are given as follows.

\begin{itemize}
  \item The Hamiltonian constraint:
\begin{align}
K_{ij}K^{ij}-\frac{1}{3}\left[2\left(\frac{N}{\beta+N}\right)^2+1\right]K^2
-\alpha_1-\alpha_2R=0.
\end{align}
 \item The momentum constraints:
\begin{align}
D_j\pi^{ij}=0,
\end{align}
where
\begin{align}
  \pi^{ij}:=
  K^{ij}-\frac{1}{3}\left(\frac{2N}{\beta+N}+1\right)K\gamma^{ij}.
\end{align}
 \item The evolution equations:
\begin{align}
&\frac{1}{N\sqrt{\gamma}}\partial_t\left(\sqrt{\gamma}\pi^{kl}\right)
\gamma_{ik}\gamma_{jl}
-\frac{1}{2}\left(\alpha_1+\frac{\alpha_3}{N}\right)\gamma_{ij}
\notag \\ &
+2\left[
K_{ik}K_j^{\;k}-\frac{1}{3}\left(\frac{2N}{\beta+N}+1\right)KK_{ij}
\right]
\notag \\ &
-\frac{1}{2}\left[
K_{kl}K^{kl}-\frac{1}{3}\left(\frac{2N}{\beta+N}+1\right)K^2
\right]\gamma_{ij}
\notag \\ &
+\left(\alpha_2+\frac{\alpha_4}{N}\right)\left(R_{ij}-\frac{1}{2}R\gamma_{ij}\right)
\notag \\ &
+\frac{\alpha_2}{N}\left(D^2N\gamma_{ij}-D_iD_jN\right)
\notag \\ &
+\frac{1}{N}\left[
D^k(\pi_{ik}N_j)+D^k(\pi_{jk}N_i)-D^k(\pi_{ij}N_k)
\right]
=0.
\end{align}
\end{itemize}

\section{Covariant form of the action}\label{app:covariantize}

In the main text,
we use the ADM decomposition and consider the action of the form~\eqref{action:Gao}
which no longer has full diffeomorphism invariance. However,
one can always restore it by introducing a St\"uckelberg field $\phi(x^\mu)$.
In this appendix, we derive the covariant form of the action
written in terms of the four-dimensional metric $g_{\mu\nu}$ and $\phi$.

The St\"uckelberg scalar field is introduced by writing the unit
normal to constant time hypersurfaces as $n_\mu=-\phi_\mu/\sqrt{2X}$,
where $\phi_\mu:=\nabla_\mu\phi$ and $X:=-\phi^\mu\phi_\mu/2$ with
$\nabla_\mu$ being the usual four-dimensional covariant derivative.
The ADM variables in the action~\eqref{action:Gao} are then replaced as follows:
\begin{align}
    N\to &\;\frac{1}{\sqrt{2X}},
    \\
    K_{ij}\to&\; {\cal K}_{\mu\nu}:=-\frac{\phi_{\mu\nu}}{\sqrt{2X}}+n_\mu a_\nu+n_\nu a_\mu 
    \notag \\ & \quad \qquad \;
    +\frac{1}{2X}n_\mu n_\nu n^\rho\nabla_\rho X,
    \\
    R\to &\; {}^{(4)}{\cal R}-{\cal K}_{\mu\nu}{\cal K}^{\mu\nu}+{\cal K}^2-2\nabla_\mu
    \left({\cal K}n^\mu-a^\mu\right),
\end{align}
where $\phi_{\mu\nu}:=\nabla_\mu\nabla_\nu\phi$, $a_\mu:=n^\rho\nabla_\rho n_\mu$,
and ${}^{(4)}{\cal R}$ is the four-dimensional Ricci scalar.
One thus obtains 
\begin{align}
    S&=\int\D^4 x\sqrt{-g}{\cal L},
    \\
    {\cal L}&=\left[\frac{1}{2}\widetilde B(\phi,X)-\widetilde f(\phi,X)\right]
    \left({\cal K}_{\mu\nu}{\cal K}^{\mu\nu}-{\cal K}^2\right)
    \notag \\ & \quad 
    +\widetilde C(\phi,X){\cal K}^2+\frac{1}{2}
    \left[\widetilde\alpha_1(\phi)+\widetilde\alpha_3(\phi)\sqrt{2X}\right]
    \notag \\ & \quad 
    +\widetilde f(\phi,X)\left[{}^{(4)}{\cal R}-2\nabla_\mu
    \left({\cal K}n^\mu-a^\mu\right)\right],
\end{align}
where
\begin{align}
    \widetilde f&=\frac{1}{2}\left[\widetilde\alpha_2(\phi)+\widetilde\alpha_4(\phi)\sqrt{2X}\right],
    \\
    \widetilde B&=\frac{\widetilde{\beta}_0(\phi)}{\widetilde{\beta}_2(\phi)\sqrt{2X}+1},
    \\
    \widetilde C&=\frac{\widetilde{\beta}_0(\phi)}{3}\cdot 
    \frac{\bigl[\widetilde\beta_1(\phi)-\widetilde\beta_2(\phi)\bigr]\sqrt{2X}}%
    {\bigl[\widetilde\beta_1(\phi)\sqrt{2X}+1\bigr]\bigl[\widetilde\beta_2(\phi)\sqrt{2X}+1\bigr]}.
\end{align}
The time-dependent functions in the action~\eqref{action:Gao} are given
by\footnote{The relation~\eqref{appeq:relation}
can be used to write the $\phi$-dependent functions
explicitly, for example, in the first
concrete model presented in Sec.~\ref{subsec:cosmo}
(with $w=0$) as
\begin{align}
\widetilde\alpha_1&=6\left\{
\left[\frac{1}{\xi}-\frac{1}{(1+\beta)^2}\right]\phi^2-\frac{h_0^2}{\xi}
\right\},
\\
\widetilde\alpha_3&=\frac{4}{\phi^2-h_0^2}\left\{
\left[\frac{1}{\xi}-\frac{1}{(1+\beta)^2}\right]\phi^2-
\left(\frac{1}{\xi}-\frac{1}{1+\beta}\right) h_0^2
\right\},
\\
\widetilde\beta_1&=-\frac{2\beta}{3(\phi^2-h_0^2)},
\\
\widetilde\alpha_2&=\widetilde\beta_0=1,\quad 
\widetilde\alpha_4=\widetilde\beta_2=0,
\end{align}
with the solution for the scalar field $\phi(t)=h_0\coth[(3/2)h_0t]$.}
\begin{align}
&\beta_0(t)=\widetilde\beta_0(\phi(t)),
\quad\beta_{1,2}(t)=\dot\phi(t)\widetilde\beta_{1,2}(\phi(t)),
\notag \\
&\alpha_{1,2}(t)=\widetilde\alpha_{1,2}(\phi(t)),
\quad
\alpha_{3,4}(t)=\dot\phi(t)\widetilde\alpha_{3,4}(\phi(t)).
\label{appeq:relation}
\end{align}

With some manipulation, the Lagrangian can be written in a more suggestive form as
\begin{align}
    {\cal L}&=\frac{\widetilde\alpha_1}{2}-\widetilde\alpha_2''(2X-X\ln X)
    +\frac{\widetilde\alpha_3}{2}\sqrt{2X}+\widetilde\alpha_4''(2X)^{3/2}
    \notag \\ & \quad 
    -\left(\frac{\widetilde\alpha_2'}{2}\ln X+2\widetilde\alpha_4'
    \sqrt{2X}\right)\Box\phi
    +\widetilde f{}^{(4)}{\cal R}+{\cal L}_{\mathrm{quad}},
\end{align}
where a prime here denotes differentiation with respect to $\phi$
and
\begin{align}
    {\cal L}_{\mathrm{quad}}&=A_1\phi_{\mu\nu}\phi^{\mu\nu}+A_2(\Box\phi)^2
    +A_3\Box\phi\phi^\mu\phi_{\mu\nu}\phi^\nu
    \notag \\ & \quad 
    +A_4\phi^\mu\phi_{\mu\rho}\phi^{\rho\nu}\phi_\nu+A_5(\phi^{\mu}\phi_{\mu\nu}\phi^\nu)^2,
\end{align}
with
\begin{align}
    &A_1=\frac{\widetilde B}{4X}-\frac{\widetilde f}{2X},\quad 
    A_2=-A_1+4X^2A_5,
    \notag \\
    &A_3=-A_4+4XA_5,
    \quad 
    A_4=\frac{\widetilde f_{,X}+A_1}{X},
    \notag \\
    &A_5=\frac{\widetilde C}{8X^3}.
\end{align}
This is a particular case of the U-degenerate theories,
i.e., higher-order scalar-tensor theories that are degenerate
when restricted to the unitary gauge~\cite{DeFelice:2018ewo}. It is easy to check that
the above Lagrangian does not satisfy the degeneracy conditions
in an arbitrary gauge if $\widetilde C\neq 0$.
In general, there is one dynamical scalar DOF in a U-degenerate theory in the unitary gauge.
However, with the above particular form of the Lagrangian, 
it turns out that $\phi$ does not propagate.

To demonstrate that $\phi$ does not propagate,
let us study the equation of motion for $\phi$ in
a Friedmann‐Lema\^{i}tre‐Robertson-Walker universe with the metric $\D s^2=-\D t^2+a^2(t)\D\Vec{x}^2$.
For simplicity, we consider the case with
$\alpha_2=\beta_0=1$ and $\alpha_4=\beta_2=0$. The equation of motion for $\phi$ then reads
\begin{align}
    \widetilde\alpha_1'\dot\phi-3H\widetilde\alpha_3\dot\phi
    -\frac{6\widetilde\beta_1\dot\phi H}{\widetilde\beta_1\dot\phi+1}{\cal P}=0,
    \label{app:eom-stuck}
\end{align}
where
\begin{align}
    {\cal P}:=
    2\frac{\D}{\D t}\left(\frac{H}{\widetilde\beta_1\dot\phi+1}\right)
    +\frac{3H^2}{\widetilde\beta_1\dot\phi+1}.
\end{align}
This equation does contain $\ddot\phi$ as well as $\dot H$. However,
the $(ij)$ components of the field equations
have the form,
\begin{align}
    {\cal P}+\frac{1}{2}\left(\widetilde\alpha_1+\widetilde\alpha_3\dot\phi\right)+\cdots=0,
\end{align}
where the ellipsis denotes the pressure of other matter fields.
The evolution equations for the metric and $\phi$ are thus degenerate
as in extended cuscuton theories~\cite{Iyonaga:2018vnu},
implying that the scalar DOF is in fact not dynamical.

Let us turn to discuss black hole solutions in the case where
\begin{align}
    &\widetilde\beta_1=\mathrm{const},\quad \widetilde\beta_2=0,
    \quad 
    \widetilde\alpha_1=\alpha_1=\mathrm{const},
    \notag \\ &
    \widetilde\alpha_2=\alpha_2=\mathrm{const},
    \quad 
    \widetilde\alpha_3=\widetilde\alpha_4=0.
\end{align}
Concerning black hole solutions,
working in the covariant action
results in much more involved equations.
However, it is not difficult to see that the field equations
admit the following configuration of the metric and $\phi$:
\begin{align}
    \D s^2&=-h(r)\D T^2+\frac{\D r^2}{h(r)}+r^2\D\Omega^2,
    \\
    \phi(T,r)&=\frac{1}{N_0}\left(T+\int^r\frac{b_0/r^2}{h\sqrt{h+b_0^2/r^4}}\D r\right),
\end{align}
where
\begin{align}
    h(r):=1+\frac{\alpha_1}{6\alpha_2}r^2-\frac{\mu_0}{r}+\frac{1-\alpha_2}{\alpha_2}\frac{b_0^2}{r^4}.
\end{align}
Note that $\phi$ is allowed to have the linear $T$ dependence
thanks to the shift symmetry $\phi\to\phi+\,$const.
This is a so-called stealth solution (when $\alpha_2=1$),
but, in contrast to
the familiar examples in the literature, we have $X\neq\,$const
in the present case.
(See Ref.~\cite{Takahashi:2020hso} for a comprehensive discussion on
stealth of solutions in quadratic degenerate higher-order scalar-tensor theories.)
Moving to the unitary gauge, $\phi(T,r)=t$, 
this solution reproduces the one obtained in Sec.~\ref{subsec:bh}.

\bibliography{refs}
\bibliographystyle{JHEP}
\end{document}